\documentclass[11pt,twoside]{article}
\usepackage{newpasp}
\usepackage{epsfig}

\usepackage{epsf}

\index{Rao, S}

\begin{document}

\title{The Cosmological Evolution of Quasar Damped Lyman-Alpha Systems}
\author{Sandhya Rao}
\affil{Dept. of Physics and Astronomy, The University of Pittsburgh, 
Pittsburgh, PA 15260}


\begin{abstract}

We present results from an efficient, non-traditional survey  
to discover damped Ly$\alpha$ (DLA) absorption-line systems with neutral 
hydrogen column densities $N_{HI}\ge2\times10^{20}$ atoms 
cm$^{-2}$ and redshifts $z<1.65$. Contrary to previous studies at
higher redshift that showed a decrease in the cosmological mass 
density of neutral gas in DLA absorbers, $\Omega_{DLA}$, with time,
our results indicate that $\Omega_{DLA}$ is consistent with remaining
constant from redshifts $z\approx4$ to $z\approx0.5$. There is no 
evidence that $\Omega_{DLA}$ is approaching the value at $z=0$. Other
interesting results from the survey are also presented. 

\end{abstract}




\section{Introduction}

Quasars, being among the most luminous and distant objects known,
are powerful probes of the Universe. Quasar absorption
lines reveal the presence of intervening gaseous clouds all way up
to redshift $z\approx 5$, back to a time when the Universe was less
than 10\% of its present age. By studying the properties of 
intervening absorption-line systems we can answer fundamental questions
about the formation and evolution of gaseous structures in the Universe. 

Intervening gas clouds (see Figure 1) are found to have neutral hydrogen 
column densities in the range $10^{12}\le N_{HI} \le 5\times10^{21}$ atoms 
cm$^{-2}$ and are thought to be associated with anything from possibly
primordial clouds in the intergalactic medium (e.g., the weakest ``Ly$\alpha$
forest'' systems) to low-luminosity galaxies, interacting systems, 
low-surface-brightness galaxies, and the evolved gaseous spheroidal and disk
components of the most luminous galaxies. 

\begin{figure}
\plotone{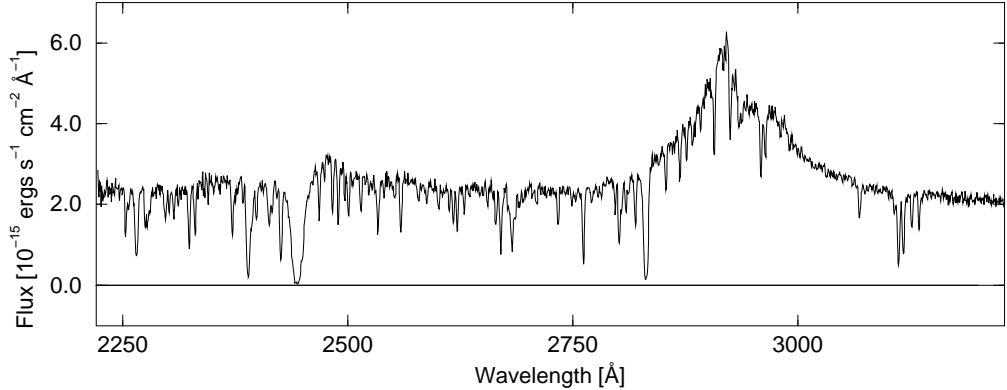}
\caption{HST-FOS spectrum of the QSO EX 0302$-$223. 
The Ly$\alpha$ emission line from the QSO is centered
at 2930\AA. Most of the absorption lines blueward of this emission line
are due to Ly$\alpha$ in intervening objects. The strongest line in this
``forest'' of lines is a ``damped Ly$\alpha$'' line at 2440\AA. }
\end{figure}

Damped Ly$\alpha$ (DLA) systems, which classically have 
$N_{HI} \ge 2\times10^{20}$ atoms cm$^{-2}$, comprise the highest column
density systems in the Ly$\alpha$ forest and contain over 90\% of the
observable HI mass in the Universe. However, these systems are very 
rare; they are 12 orders of magnitude less abundant than the weakest
systems in the forest (e.g., Hu et al. 1995, AJ, 110, 1526). In the 
past, ground-based surveys for DLA absorption have been used to study 
the distribution of HI at high redshift (e.g., Wolfe et al. 1995, ApJ, 454, 698). 
Since the Ly$\alpha$ line falls in the UV for redshifts $z<1.65$, valuable
HST time is required to discover DLA systems at low redshift. With
$z<1.65$ spanning the most recent 77\% of the age of the Universe
($\Lambda=0$ and $q_0=0.5$), the determination of the statistics and 
properties of DLA systems at these redshifts is crucial for understanding
the evolution of galaxies. Here, we present results from an efficient, 
non-traditional survey to discover DLA absorption-line systems at
redshifts $z<1.65$.

\section{The Survey}

For a detailed description of the survey the reader is referred to
Rao \& Turnshek (2000, ApJS, in press, astro-ph/9909164).
 Here, we list some salient points.

\noindent
1. All DLA systems exhibit low-ionization metal line transitions.
Thus, MgII can be used as a tracer for DLA absorption.

\noindent
2. Since the statistics of low-redshift ($0.1<z<2.2$) MgII systems
are known (Steidel \& Sargent 1992, ApJS, 80, 1), the fraction of DLA 
systems in a MgII selected sample can be used to derive
 the DLA incidence, $n_{DLA}(z)$.

\noindent
3. The measured column densities of the DLA lines give the DLA
cosmological mass density, $\Omega_{DLA}(z)$, and their column
density distribution, $f(z,N)$.

\noindent
4. DLA lines can be searched for in over 250 known, $z<1.65$, 
MgII systems. In total, from our IUE and HST surveys (Rao, Turnshek,
\& Briggs 1995, ApJ, 449, 488; Rao \& Turnshek 2000), we now have Ly$\alpha$
information on 87 MgII systems with rest equivalent width
$W_0^{\lambda2796} \ge 0.3$\AA\ and redshifts $0.11<z<1.65$.

\section{The Results}

We uncovered 14 DLA systems in the survey, more than
doubling the number of known low-redshift DLA systems. 
Again, the reader is referred to Rao \& Turnshek (2000) for details on 
each system. The significant
results from the survey are shown in Figures 2, 3, 4, and 5, and are
discussed in the figure captions.

\begin{figure}
\begin{center}
\epsfig{file=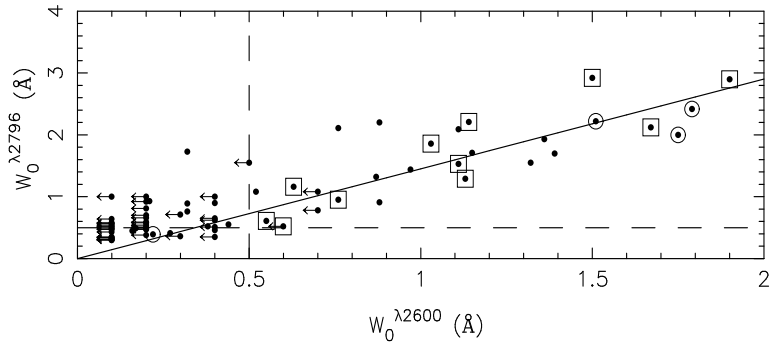,height=1.8in,width=4.0in}
\end{center}
\caption{MgII $\lambda 2796$ rest equivalent width, $W_0^{\lambda2796}$,
vs. FeII $W_0^{\lambda2600}$. The DLA systems in our sample are marked with
open squares. The open circles represent previously known 21 cm 
absorbers that were excluded from our unbiased MgII sample. Arrows
are upper limits. Approximately 50\% of the systems with $W_0^{\lambda2796}
> 0.5$ \AA\ and $W_0^{\lambda2600} > 0.5$ \AA\ (demarcated by the dashed
lines) have DLA absorption. All systems in this regime have 
$N_{HI} > 10^{19}$ atoms cm$^{-2}$. The two rest equivalent widths
are linearly correlated with slope 1.45.}
\end{figure}

\begin{figure}
\begin{center}
\epsfig{file=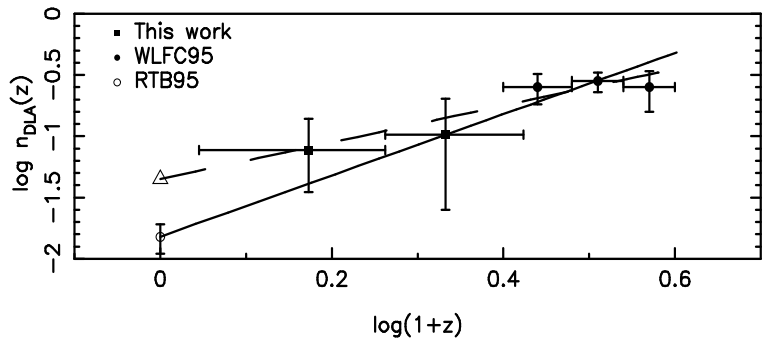,height=1.8in,width=4.0in}
\end{center}
\caption{Evolution of the incidence of DLA systems, where $n_{DLA}(z)$
is the number of DLA systems per unit redshift along a line of sight.
The high-redshift points are from Wolfe et al. (1995) and the $z=0$
point is derived from the observed HI distribution in local spiral
galaxies (Rao, Turnshek, \& Briggs 1995). The lines are power laws
of the form $n_{DLA}(z) = n_0 (1+z)^\gamma$. For no intrinsic evolution
of the absorbers, $\gamma=0.5$ for $q_0=0.5$ and $\gamma=1$ for $q_0=0$.
The solid line has $\gamma=2.5$. The dashed line has $\gamma=1.5$ and 
excludes the $z=0$ data point, but is extrapolated to $z=0$ (open triangle).
Thus, there is strong evidence for evolution in the incidence of DLA 
systems if the $z=0$ point is included.
}
\end{figure}

\begin{figure}
\begin{center}
\epsfig{file=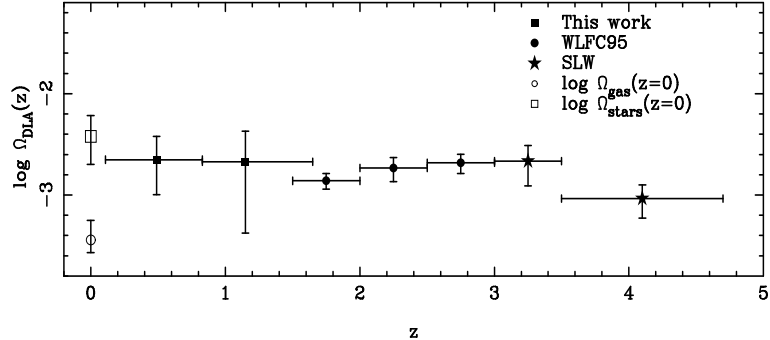,height=1.8in,width=4.0in}
\end{center}
\caption{The cosmological mass density of neutral gas in DLA systems,
$\Omega_{DLA}$, as a function of redshift ($\Lambda=0$, $H_0=65$ km s$^{-1}$
Mpc$^{-1}$, and $q_0=0.5$). Contrary to studies at high redshift (Wolfe et al.
1995; Storrie-Lombardi \& Wolfe 2000, astro-ph/0006044) which conclude that 
$\Omega_{DLA}$
decreases with time for $3.5>z>1.5$, our results indicate that $\Omega_{DLA}$
is consistent with remaining constant from redshifts $z\approx4$ to 
$z\approx0.5$. 
There is no evidence that $\Omega_{DLA}$ is approaching $\Omega_{gas}(z=0)$.
The constant value for $\Omega_{DLA}$ might indicate that DLA absorption
lines track a slowly evolving population of objects. This is also 
consistent with the lack of evolution seen in their metallicities
(Pettini et al. 1999, ApJ, 510, 576).
}
\end{figure}

\begin{figure}
\begin{center}
\epsfig{file=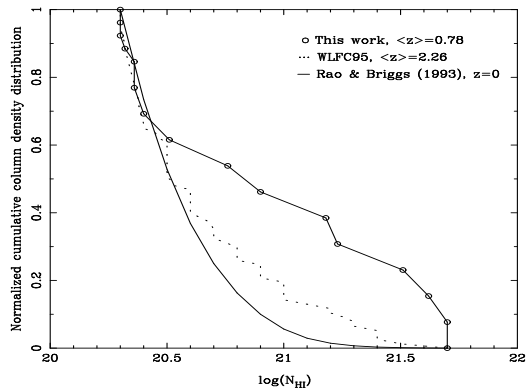,height=2.0in,width=2.7in}
\end{center}
\caption{The normalized cumulative column density distribution for the
low-redshift DLA sample, the Wolfe et al. (1995) high-redshift DLA sample,
and local galaxies (Rao \& Briggs 1993, ApJ, 419, 515). The low-redshift 
DLA sample clearly
contains a much larger fraction of high column density systems than either
of the other two samples. The $N_{HI}^{-3}$ fall-off in the column
density distribution of spiral disks results in the characteristic shape of
the solid curve. The low-redshift sample has a distribution that is 
different at the 99.99\% confidence level indicating that low-redshift
DLA absorption lines do not arise solely in galactic disks (see also
Turnshek et al., these proceedings). 
}
\end{figure}

\end{document}